\documentclass{article}
\usepackage[utf8]{inputenc}
\usepackage{graphicx}
\usepackage{caption}
\usepackage{subcaption}
\usepackage{amsmath}
\usepackage{authblk}

\title{Self similar Shocks in Atmospheric Mass Loss due to  Planetary Collisions}
\author[1]{Almog Yalinewich}
\author[1]{Andrey Remorov}
\affil[1]{Canadian Institute for Theoretical Astrophysics, 60 St. George St., Toronto, ON M5S 3H8, Canada}
\date{\today}

\begin{document}

\maketitle

\begin{abstract}
    We present a mathematical model for the propagation of the shock waves that occur during planetary collisions. Such collisions are thought to occur during the formation of terrestrial planets, and they have the potential to erode the planet's atmosphere. We show that under certain assumptions, this evolution of the shock wave can be determined using the method of self similar solutions. This self similar solution is of type II, which means that it only applies to a finite region behind the shock front. This region is bounded by the shock front and the sonic point. Energy and matter continuously flow through the sonic point, so that energy in the self similar region is not conserved, as is the case for type I solutions. Instead, the evolution of the shock wave is determined by boundary conditions at the shock front and at the sonic point. We show how the evolution can be determined for different equations of state, allowing these results to be readily used to calculate the atmospheric mass loss from planetary cores made of different materials.
\end{abstract}

\section{Introduction}
The formation of terrestrial planets is thought to proceed in two phases. In the first phase, rocky cores form by accretion in the protoplanetary disc \cite{Armitage2018AFormation}, and in the second stage, multiple planetary cores merge to form planets after the disc evaporates \cite{Agnor1999OnFormation, Chambers2001MakingPlanets}.  

Theoretical models predict that when the disc evaporates, it will leave behind a large number of tightly packed planetary cores. Dynamical instabilities will excite eccentricities which will cause the orbits to intersect and the core to collide \cite{Chambers1996TheSystems, Zhou2007Post-OligarchicSystems, Obertas2017TheStar, Rice2018SurvivalEncounter}. Among other things, these collisions can erode a planetary atmosphere. 

It is difficult to observe planetary collisions directly, as they are expected, in the most optimistic scenario, to produce a faint, short (few hours) and hard (X-ray and EUV) transient \cite{Zhang2003ElectromagneticCollisions}. Study of this phase is therefore based on theoretical or computational models, and indirect observational evidence, such as isotope ratios on Earth \cite{Schlichting2018AtmosphereLosses} or debris rings around exoplanetary systems \cite{Thompson2019StudyingSOFIA}. Planetary collisions have been suggested as possible mechanism to explain systems like Kepler 107 \cite{Bonomo2019ASystem}, where a pair of neighbouring planets have very different average density. Kepler 107b has a low density, indicating the presence of an atmosphere, while Kepler 107c, while farther away from the host star, has a density almost three times larger than 107b, indicating a bare rocky core devoid of an atmosphere. If both planets started out with a thick atmosphere, other mechanisms, such as photoevaporation, could not have stripped Kepler 107c of its atmosphere while sparing 107b.

Simulations of giant collisions have been carried out by many authors in the past, e.g. \cite{Stewart2011CollisionsFormation}. However, these authors mostly considered the change in solid mass. For this purpose, authors ran many simulations, spanning a wide range of parameters. In contrast, for simulations of atmospheric mass loss due to giant collisions, most works focused on a single specific case, e.g. \cite{Liu2015GIANTSUPER-EARTHS} which considered Kepler 36.

A systematic study of atmospheric mass loss for head on collisions has been carried out in \cite{Schlichting2015AtmosphericImpacts, Yalinewich2018AtmosphericImpacts}. The model they developed consists of two phases. In the first phase a shock wave emerges from the impact site and propagates through the planet. The second stage focuses on the interface between the ground and the atmospheric, and notice that when the ground moves due to the shock propagating through the target, another shock wave emerges from the gas - ground interface and moves through the atmosphere. Since the atmosphere has a declining density profile, the shock waves that travels through it accelerates as it moves up. At some point the shock wave exceeds the escape velocity, and all the gas above this point is lost. For this reason, some of the gas column above the ground is expelled even if the ground move slower than the escape velocity.

Propagation of a shock due to an impact event has been studied using simulations \cite{Potter2012ConstrainingImpact, Monteux2019ShockRevisited} and experiments \cite{Holsapple2012MomentumScaling, Mazzariol2019ExperimentalMaterials, Holsapple1987ThePhenomena}. However, under certain assumptions, the shock can be described analytically, using the theory of self similar solutions \cite{Barenblatt1996ScalingAsymptotics}. The most famous solution of this kind is the Sedov Taylor solution for the strong adiabatic explosion \cite{Sedov1946PropagationWaves, Taylor1950TheDiscussion}. In this case, the motion of the shock front is determined using conservation of energy. Such cases where the self similar motion can be determined using conservation laws is called type I solutions. In contrast, the shock waves in planetary collisions are type II solution. In such cases, conservation laws cannot be used and the motion is determined by the behaviour near a singularity \cite{Waxman1993SecondtypeProblem}.

The intention of this paper is to explain how the evolution of the shock in the interior can be described using the theory of self similarity. The structure of this paper is as follows: in section \ref{sec:ipp} we describe the self similar solution of the shock propagating in the core. In section \ref{sec:aml} we demonstrate how the atmospheric mass loss can be estimated. Finally. we discuss the results in section \ref{sec:conclusion}.

\section{Core Shock} \label{sec:ipp}

\subsection{Impulsive Piston Problem}

Let us consider an impact on a very large planet, so that the radius of curvature is unimportant and the ground can be considered flat. Such an impact will launch a shock wave that spreads as a hemisphere below the ground. When the radius of the shock wave is much larger than the size of the impactor, there are no other length scale in this problem and the shock radius grows as some power law in time $R \propto t^{\alpha}$. The purpose of this section is to estimate $\alpha$.

Even if we assume the shock to have azimuthal symmetry, the equations of motion still depend on two spatial coordinates and time. We therefore turn to an even simpler problem - the impulsive piston problem \cite{Adamskii1959TheLaw, Zeldovich1967PhysicsPhenomena}. In this case, we consider a slab symmetric, one dimensional analogue to the self similar impact problem. In this scenario, a thin wafer hits a much thicker slab of material and both are perfectly cold prior to the collision. As a result of this collision a shock wave emerges from the contact surface and travels into the target. Once the shock wave has travelled a distance much larger than the width of the wafer, there is no other relevant length scale in this process other than the position of the shock front, and hence we expect the shock wave to evolve in a self similar way. 

What's incredible about the impulsive piston problem is that the relation between the shock velocity and the swept up mass $v \propto m^{\delta}$ holds also in the three dimensional impact problem \cite{Yalinewich2018AtmosphericImpacts}, i.e. a power law with the same exponent $\delta$. In the remainder of this section we present the mathematical formulation of the impulsive piston problem.

\subsection{Self Similar Equations}
The slab symmetric hydrodynamic equations in one dimension are given by \cite{Landau1987FluidMechanics}

\begin{equation}
\frac{\partial \rho }{\partial t}+ 
\rho \frac{\partial v }{\partial x} + v \frac{\partial \rho}{\partial x} = 0 \label{eq:mass_conservation}
\end{equation}

\begin{equation}
\frac{\partial v}{\partial t} +  v\frac{\partial v}{\partial x}+ 
\frac{1}{\rho} \frac{\partial p}{\partial x}  = 0 \label{eq:momentum_conservation}
\end{equation}

\begin{equation}
\frac{\partial s}{\partial t} + v \frac{\partial s}{\partial x}= 0 \label{eq:entropy_conservation}
\end{equation}
We assume an ideal gas equation of state
\begin{equation}
s = \ln p - \gamma \ln \rho
\end{equation}
Naturally, one would expect the material in a planetary core to have a more complicated equation of state. However, we argue that at very high pressures and temperatures all materials behave like ideal gases.
We can replace the pressure by the sound speed using the relation of $p =  \rho c^2 / \gamma$. After this substitution we have a system of partial differential equations in $v, c, \rho$. We make the assumption that one dimensional shock has a self similar behaviour, so the position of the shock front $X(t)$ evolves as a power law in time
\begin{equation}
    X(t) \propto t^\alpha \, . \label{eq:shock_power_law}
\end{equation}
where $\alpha$ is some number which will determined later. We can use the conservation of energy or momentum to obtain limits on this parameter $\frac{2}{3}>\alpha>\frac{1}{2}$ \cite{Zeldovich1967PhysicsPhenomena}. The partial differential equations can be reduced to ordinary differential equations in the dimensionless position  $\chi = x/X$. We define dimensionless hydrodynamical variables $V, C, D$ in the following way
\begin{equation}
    v(x,t) = \frac{d X(t)}{dt}\chi  V  \quad \quad
    c(x,t) = \frac{d X(t)}{dt} \chi C \quad \quad
    \rho(x,t) = \rho_0 D \label{eq:self_similar_vars}
\end{equation}
To get rid of $\frac{d^2 X(t)}{dt^2}$ terms, we make the substitution: $\frac{d^2 X(t)}{dt^2} = \left(\frac{d X(t)}{dt}\right)^2\frac{\delta}{X(t)}$. Substituting equation \ref{eq:shock_power_law} into this gives a relation between $\delta$ and $\alpha$: $\delta = 1 - \frac{1}{\alpha}$. The energy and momentum conservation limits on this parameter are $-\frac{1}{2} > \delta > -1$. After making these adjustments and simplifying, the hydrodynamic equations equations \ref{eq:mass_conservation}, \ref{eq:momentum_conservation} and \ref{eq:entropy_conservation} become

\begin{equation}
D (V + \chi V') + V \chi D' - \chi D' = 0
\end{equation}

\begin{equation} 
C (2 C D + C \chi D' + 2 D \chi C') + D V \delta \gamma + D \gamma (V (V + \chi V') - V - \chi V') = 0
\end{equation}

\begin{equation}
2 C D + C \chi \gamma D'  - V (- 2 C D + C \chi \gamma D' - C \chi D' - 2 D \chi C')  - (2 C D + C \chi D' + 2 D \chi C') = 0
\end{equation}

where a prime denotes derivation with respect to $\chi$. These equations are now linear in $V', C', D'$ so these can be isolated. The relevant equations are $V', C'$, and are given by

\begin{equation}
V' = \frac{- C^{2} V \gamma - 2 C^{2} \delta + V^{3} \gamma + V^{2} \delta \gamma - 2 V^{2} \gamma - V \delta \gamma + V \gamma}{\chi \gamma \left(C^{2} - V^{2} + 2 V - 1\right)} \label{eq:dVdx}
\end{equation}

\begin{equation}
C' = \frac{N_1}{D_1}
\end{equation}
where
\begin{equation*}
N_1 = C \left(V \gamma \left(V - 1\right) \left(V \gamma - V - \gamma + 1\right) - 2 \left(C^{2} - \gamma \left(V - 1\right)^{2}\right) \left(V + \delta - 1\right) - \right.
\end{equation*}
\begin{equation}
 \left. - \left(2 C^{2} + V^{2} \gamma + V \delta \gamma - V \gamma\right) \left(V \gamma - V - \gamma + 1\right)\right)
\end{equation}
\begin{equation}
D_1 = 2 \chi \left(C^{2} \left(V - 1\right) + C^{2} \left(V \gamma - V - \gamma + 1\right) - \gamma \left(V - 1\right)^{3}\right)
\end{equation}

Dividing one equation by the other, we get a single ODE

\begin{equation}
\frac{dC}{d V} = \frac{N_2}{D_2} \label{eq:dCdV_raw}
\end{equation}
where:
\begin{equation*}
    N_2 = C \gamma \left(- V \gamma \left(V - 1\right) \left(V \gamma - V - \gamma + 1\right) + 2 \left(C^{2} - \gamma \left(V - 1\right)^{2}\right) \left(V + \delta - 1\right) + \right.
\end{equation*}
\begin{equation}
    \left. + \left(2 C^{2} + V^{2} \gamma + V \delta \gamma - V \gamma\right) \left(V \gamma - V - \gamma + 1\right)\right) \left(C^{2} - V^{2} + 2 V - 1\right)  \label{eq:dCdV_num}
\end{equation}

and
\begin{equation}
D_2 = 2 \left(C^{2} \left(V - 1\right) + C^{2} \left(V \gamma - V - \gamma + 1\right) - \gamma \left(V - 1\right)^{3}\right) \times  \label{eq:dCdV_den}
\end{equation}
\begin{equation*}
    \times \left(C^{2} V \gamma + 2 C^{2} \delta - V^{3} \gamma - V^{2} \delta \gamma + 2 V^{2} \gamma + V \delta \gamma - V \gamma\right) \, .
\end{equation*}

\subsection{Boundary Conditions}
In order integrate equation \ref{eq:dCdV_raw}, boundary conditions and the value of $\alpha$ are needed. The boundary conditions at the shock front are given by the Rankine Hugoniot conditions for a strong shock \cite{Landau1987FluidMechanics, Zeldovich1967PhysicsPhenomena}
\begin{equation}
    V_f = \frac{2}{\gamma+1} \label{eq:v_shock_front}
\end{equation}
and
\begin{equation}
    C_f = \frac{\sqrt{2 \gamma \left(\gamma-1\right)}}{\gamma+1} \, .  \label{eq:C_final}
\end{equation}
In some self similar problems, like the Sedov Taylor explosion, the parameter $\alpha$ can be be inferred directly from conservation laws. Such problems are known as type I solutions. In our case, however, $\alpha$ is determined by the condition that the hydrodynamic trajectory passes smoothly through some singularity. These are called type II solutions \cite{Waxman1993SecondtypeProblem}. In such cases the self similar solution only applies to a finite portion of space, bounded by the shock front and the singularity. This singularity occurs at a point where information cannot propagate back to the shock front, and is therefore also referred to as the sonic point. Matter, momentum and energy continuously flow through the sonic point, so that they are not conserved in the self similar region.

Equation \ref{eq:dVdx} has a singularity when 
\begin{equation}
C = 1- V \label{eq:sonic_line}
\end{equation}
We call this curve the sonic line. On the sonic line, the denominator (equation \ref{eq:dCdV_den}) vanishes. To prevent a divergence, the numerator also has to vanish. This happens on a specific point on the sonic line, which we call the sonic point

\begin{equation}
V_{s} = \frac{2}{2 - \gamma}
\end{equation}
and
\begin{equation}
C_{s} = - \frac{\gamma}{2-\gamma}
\end{equation}

Knowing the boundary conditions and $\frac{d C}{d V}$, we can integrate from the sonic point to the shock front, as shown in Figure 1. The starting point is slightly shifted from the sonic point in the positive $V$ direction by $dV_i$, and in $C$ direction by $d V_i \frac{d C}{d V}\Bigr|_{sonic}$ for some small $d V_i \ll 1$. The slope at the sonic point is given by the equation: 
\begin{equation}
    \frac{d C}{d V}\Bigr|_{sonic} = \frac{\gamma}{8}+\frac{1}{2} + \frac{\gamma^{2} + \left(2-\gamma\right) \sqrt{\gamma \left(9 \delta^{2} \gamma - 16 \delta^{2} - 12 \delta \gamma + 8 \delta + 4 \gamma\right)}}{8 \left(\delta \gamma - 2 \delta - \gamma\right)}
\end{equation}

\begin{figure}
\centering
\begin{subfigure}{.8\textwidth}
  \centering
  \includegraphics[width=1\linewidth]{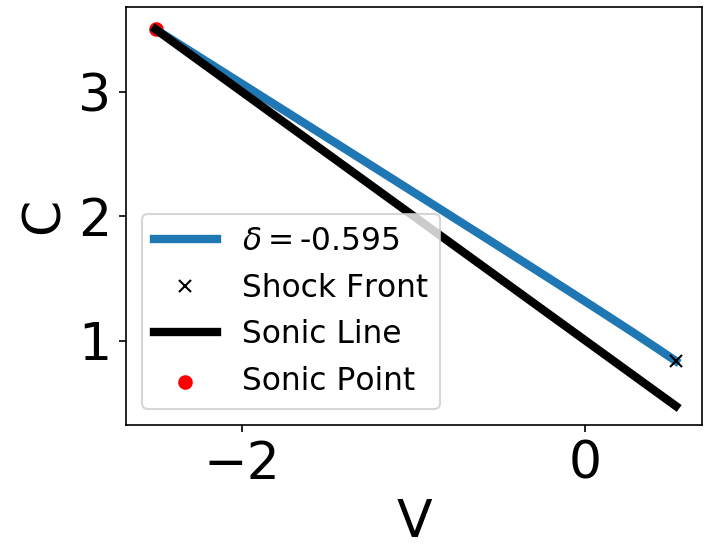}
\end{subfigure}
\begin{subfigure}{.8\textwidth}
  \centering
  \includegraphics[width=1\linewidth]{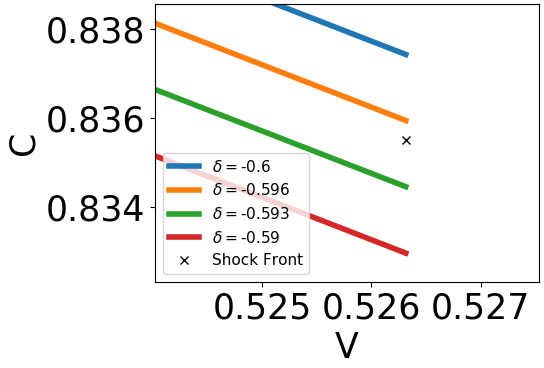}
\end{subfigure}
\caption{Numerically integrated Hydrodynamic trajectories  for $\gamma = 2.8$, as an example for the general behaviour in the case $\gamma > 2$, for various values of $\delta$. The entire domain is shown in the top panel, and a zoomed in plot at shock front point is shown in the bottom panel. The integration proceeds continuously from the sonic point (red circle) to the shock front (black cross). \label{fig:gamma_gt_2}
}
\label{fig:fig2}
\end{figure}

The correct value of $\delta$ is such that the curve $C\left(V\right)$ satisfies both boundary conditions at the sonic point and the shock front. In practice, we use the shooting method to find the value of $\delta$. We guess a value for $\delta$, numerically integrate equation \ref{eq:dCdV_raw} w.r.t to $V$ from the sonic point to the shock front, and note the value of $C$ at the shock front. We then use the bisection method to refine the value of $\delta$ to minimise the distance between the value of $C$ obtained from numeric integration and its theoretical value (equation \ref{eq:C_final}). An example for some hydrodynamic $C\left(V\right)$ trajectories for the same value of $\gamma$ but different values of $\delta$ can be seen in figure \ref{fig:gamma_gt_2}, for $\gamma = 2.8$.

\begin{figure}
\centering
\begin{subfigure}{.7\textwidth}
  \centering
  \includegraphics[width=1\linewidth]{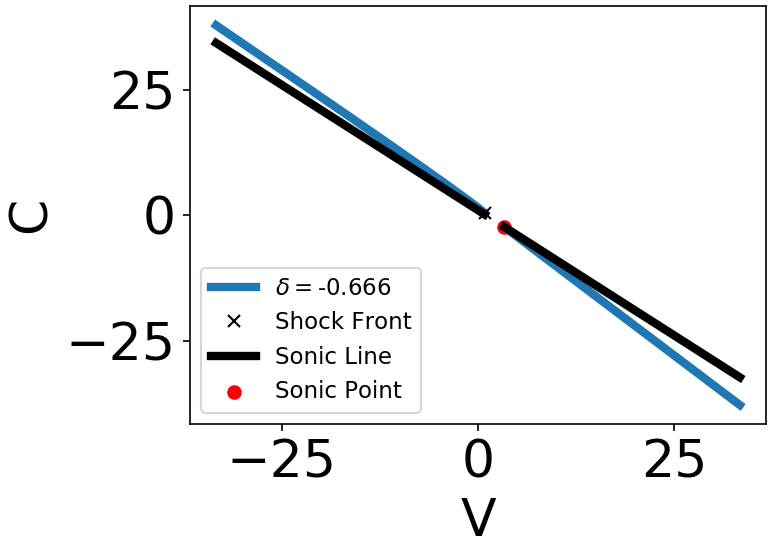}
\end{subfigure}
\begin{subfigure}{.7\textwidth}
  \centering
  \includegraphics[width=1\linewidth]{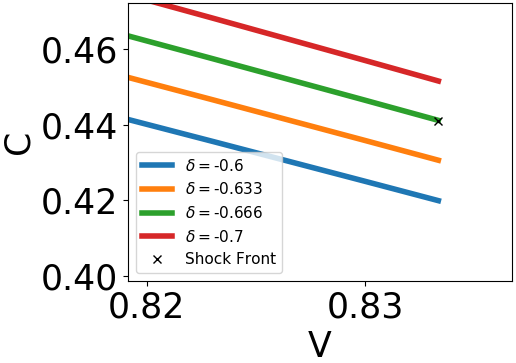}
\end{subfigure}
\caption{Numerically integrated Hydrodynamic trajectories, integrated from the sonic point (red dot) to the shock front (black cross) for $\gamma = 7/5$, as an example for the case $2 > \gamma > 1$, for different values of $\delta$. The entire domain is shown in the top panel, and a zoomed in plot at shock front point is shown in the bottom panel. The integration goes from the sonic point in the fourth quadrant and proceeds down and to the right. After passing through $\chi = 0$ the integration reappears in the second quadrant (top left corner) and travels to the shock front point.
}
\label{fig:gamma_lt_2}
\end{figure}

In the case $\gamma>2$, numerical integration is straightforward since both the sonic point and the shock front are on the same side of the impact site (i.e. both $x>0$ when the impact site is $x=0$). The case $2 > \gamma > 1$ is more complicated because the integration goes through $x=0$. Because of the way we defined the self similar variables (equation \ref{eq:self_similar_vars}), they diverge at $x=0$, even though the physical quantities $v,c$ remain finite. We can circumvent this difficulty by noting that the Mach number, i.e. the ratio between the velocity and the speed of sound, remains finite and changes continuously across $x=0$. The sonic point in the case $2>\gamma>1$ lies in the fourth quadrant of the of the $C-V$ plane (i.e. positive $V$ and negative $C$). Numerical integration proceeds to even higher values of $V$, moving away from the shock front. Far away from the sonic point, the curve attains some asymptotic slope, and this slope is the Mach number at $x=0$, which we'll denote by $\mathcal{M}_0$. Once we have this piece of information, we can restart the numerical integration in the second quadrant ($C$ is positive and $V$ is negative), beginning from some arbitrarily highly negative $V=V_r$ (subscript r for restart) and $C_r = V_r/\mathcal{M}_0$. From this point we can continue the numerical integration all the way to the shock front. An example for several hydrodynamic trajectories for $\gamma = 7/5$ can be seen in figure \ref{fig:gamma_lt_2}.

\begin{figure}
  \centering
  \includegraphics[width=0.8\linewidth]{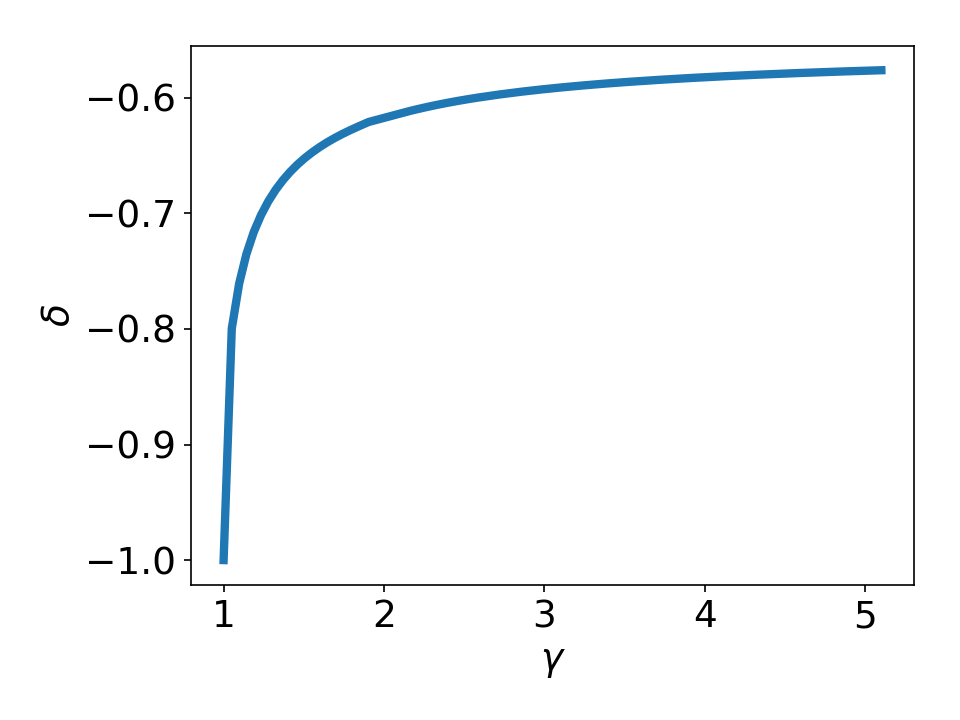}
  \caption{Shock deceleration parameter $\delta$ as a function of the adiabatic index $\gamma$ for the impulsive piston problem.}
\label{fig:delta_vs_gamma}
\end{figure}

Thus, for every value of $\gamma$ we can obtain the corresponding value of $\delta$. The relation between them is shown in figure \ref{fig:delta_vs_gamma}. These results are consistent with previous works \cite{Adamskii1959TheLaw}. We note that in the limit $\gamma = 1$, we get $\delta = -1$, which corresponds to momentum conservation. However, in the limit $\gamma \rightarrow \infty$, the solution does not converge to $\delta=-1/2$, which corresponds to energy conservation, but to a different value. More on this in the next section.

We note that the results obtained in this section are in accord with the numerical simulations presented in \cite{Yalinewich2018AtmosphericImpacts}, which found that for the three dimensional impact problem, the shock decelerates roughly as $\dot{X} \propto m^{-2/3}$ (where $m$ is the swept up mass) for $\gamma = 5/3$, where the model presented here predicts $\delta = -0.64$. The model here also explains impact experiments, although the translation between the theoretical results and the experimental results is not straightforward. In experiments, like those described in \cite{Holsapple1993TheSciences}, a projectile is fired at a slab of some material and the radius of the resulting crater is measured. In materials without shear strength, like fluids, the crater basin stops growing when the pressure inside the crater is comparable to the hydrostatic pressure. The analysis above gives us a relation between the shock velocity and the swept up mass $\dot{R} \propto m^{\delta}$, where $m$ is the swept up mass, and in a three dimensional case it scales as a cube of the radius $m \approx \rho R^3$, where, for simplicity, $\rho$ is the initial mass density of both projectile and target. If we denote the radius of the projectile by $a$, and we assume that the shock sweeps a mass comparable to the projectile before decelerating, then the relation between the shock velocity and the crater radius is roughly given by
\begin{equation}
    \dot{R} \approx U \left(\frac{R}{a}\right)^{3 \delta}
\end{equation}
where $U$ is the initial projectile velocity. The pressure inside the crater is roughly given by $\rho \dot{R}^2$, while the hydrostatic pressure is given by $\rho g R$, where $g$ is the gravitational acceleration. Equating the two pressures yields a relation between the volume of the crater $R_c^3$ (where $R_c$ is the terminal size of the crater) and the properties of the projectile
\begin{equation}
    \pi_v = \frac{R_c^3}{a^3} = \left(\frac{g a}{U^2}\right)^{\frac{3}{6 \delta - 1}} = \pi_1^{\frac{3}{6 \delta - 1}}
\end{equation}
where we define the same dimensionless variables $\pi_v = R_c^3/a$, $\pi_2 = g a/U^2$ as \cite{Holsapple1993TheSciences}. To produce a prediction, we need to estimate the effective adiabatic index. This can be done in different kinds of high pressure experiments, where a material is shocked and the material $U_m$ and shock velocity $U_s$ are measured. At high velocities, the ratio between the two tends to a constant, $U_s/U_m = \beta$, and for an ideal gas $\beta = \left(\gamma+1\right)/2$, so $\gamma = 2 \beta - 1$. For water, equation of state experiments yields $\beta = 1.78$ \cite{Gojani2016ShockGelatin}, so the effective adiabatic index $\gamma = 2.56$ our model yields $\delta = -0.61$ and $d \ln \pi_v / \ln \pi_2 = -0.643$. The value obtained from experiments in water is $d \ln \pi_v / d \ln \pi_2 = -0.648$ \cite{Holsapple1993TheSciences}. Overall, despite our crude approximations, we obtain a value relatively close to the experimental value.

\subsection{Asymptotic Case}

In this section we consider the case $\gamma \rightarrow \infty$. In this case we cannot use the equations developed above. Part of the reason is that the integration domain $V \in \left[-\frac{2}{\gamma-2}, \frac{2}{\gamma+1}\right]$ shrinks to zero width in this limit. To overcome this difficulty, we define a new variable $W = \gamma V$. In this new variable, the integration domain becomes $W \in \left[-2,2\right]$, and so remains finite in the limit $\gamma\rightarrow \infty$. The expression for the derivative in the new variable is given by

\begin{equation}
    \frac{d C}{d W} = \frac{C \left(2 C^{2} + W \delta + 2 \delta - 2\right)}{2 \left(C^{2} W + 2 C^{2} \delta + W \delta - W\right)} \, .
\end{equation}
At the sonic point, the slope is given by

\begin{equation}
    \left. \frac{d C}{d W} \right|_s = \frac{\delta}{4 \left(\delta-1\right)} \, .
\end{equation}

\begin{figure}
    \centering
    \includegraphics[width=0.9\linewidth]{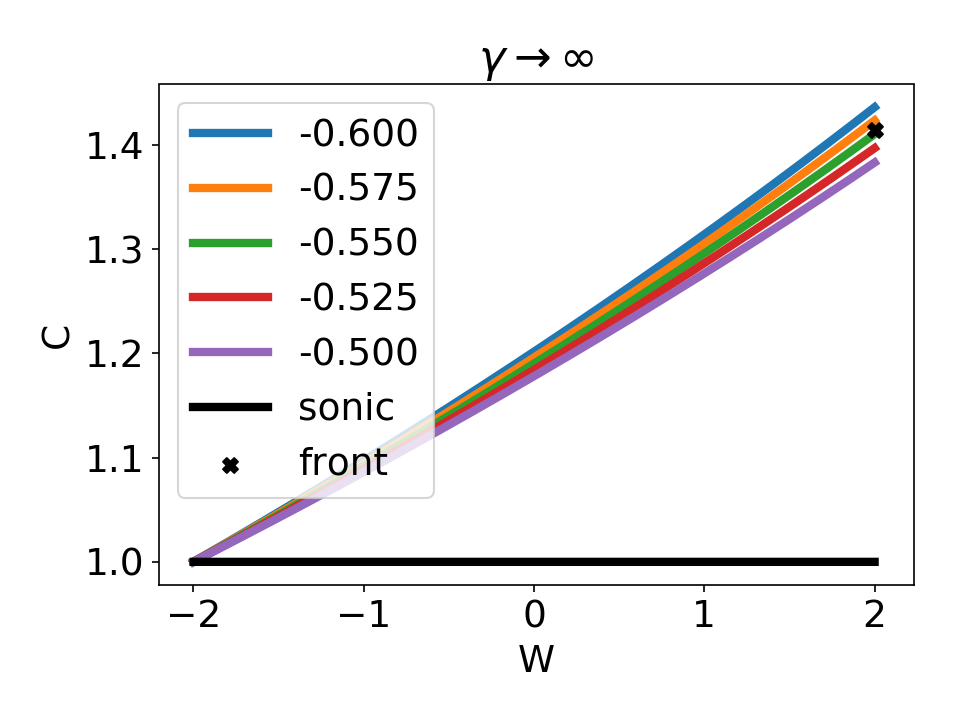}
    \caption{Hydrodynamic trajectories for the case $\gamma \rightarrow \infty$. Different coloured lined represent trajectories with different values of $\delta$. The sonic line is represented by the black line, and the shock front by the black cross.}
    \label{fig:gamma_infinity}
\end{figure}

We can determine $\delta$ using the shooting method described in the previous section. This case is actually simpler, because it does not depend on $\gamma$, so $C=1$ at the sonic point and $C = \sqrt{2}$ at the shock front. A few hydrodynamic trajectories for different values of $\delta$ are shown in figure \ref{fig:gamma_infinity}. Using numerical root finding methods we find that $\lim_{\gamma\rightarrow \infty}\delta \approx -0.557$, in accordance with \cite{Adamskii1959TheLaw}. Interestingly, this result is different from the energy conservation limit $\delta = -1/2$. 

\subsection{Analytic Case}

For most values of $\gamma$, numerical integration has to be performed to obtain the power law index $\delta$. However, for $\gamma = \frac{7}{5}$, it is possible to obtain a completely analytic solution \cite{Adamskii1959TheLaw}. Here we will briefly describe the analytic solution and the hydrodynamic profiles in the case $\gamma = \frac{7}{5}$. The dimensionless hydrodynamic quantities are

\begin{equation}
    D = 6 (5-4\chi)^{-5/2}
\end{equation}

\begin{equation}
    V = \frac{5}{6} (2-\frac{1}{\chi})
\end{equation}

\begin{equation}
    C = \frac{\sqrt{7}}{6} \frac{\sqrt{5 - 4\chi}}{\chi}
\end{equation}

We can see that at the shock front, where $\chi = 1$, the dimensionless velocity and speed of sound are $V = \frac{5}{6}$, $C = \frac{\sqrt{7}}{6}$, as verified by the coordinates of the black X in second graph of Figure 2. We also note that $V, C$ diverge as $\chi \rightarrow 0$. From the shooting method, we obtain $\delta = -\frac{2}{3}$, which is consistent with the results in Figure 2. 

\section{Atmospheric Mass Loss} \label{sec:aml}

In this section we demonstrate how the atmospheric mass loss from a giant collision is estimated. The dependence of the mass loss on impactor size and velocity in head on collisions was studied in \cite{Yalinewich2018AtmosphericImpacts}, so in this section we consider the effects of obliquity. For this purpose, we run simulations of collisions between planetary cores and planets. These simulations were run using the moving mesh numerical simulation RICH \cite{Yalinewich2015Rich:Mesh}. We use an ideal gas equation of state with an adiabatic index $\gamma = 5/3$. The planet and planetary core are described as blobs with uniform density. Since the simulation cannot handle actual vacuum, we fill the computational domain with a very tenuous gas, whose density is lower than that of the planet by a factor of $10^9$. We do not include gravity, and the entire domain has an initially uniform pressure such that the speed of sound in the planet is $10^{-6}$ of the impact velocity. Lastly, to simplify the calculation, we perform these simulations in a 2D Cartesian geometry. This means that the planet and planetary cores are not represented by spheres, but by cylinders. As opposed to the other simplifications, this last one cannot be justified, and might lead to a large discrepancy with other results. The only reason we employ this assumption is that it simplifies the analysis greatly, and since the purpose of this section is mostly pedagogical.

\begin{figure}
\centering
\begin{subfigure}{.49\textwidth}
  \centering
  \includegraphics[width=1\linewidth]{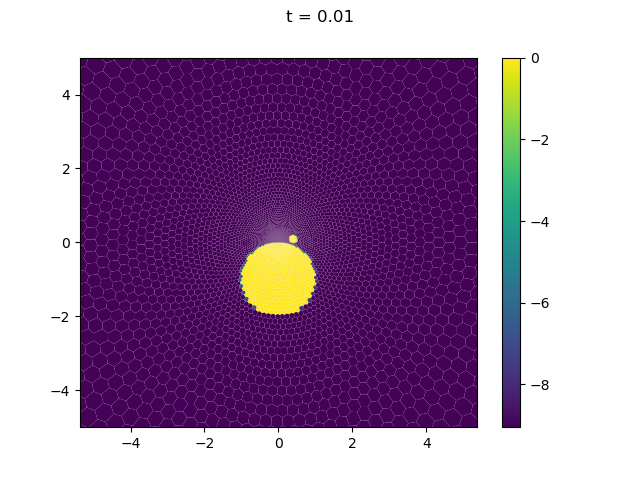}
\end{subfigure}
\begin{subfigure}{.49\textwidth}
  \centering
  \includegraphics[width=1\linewidth]{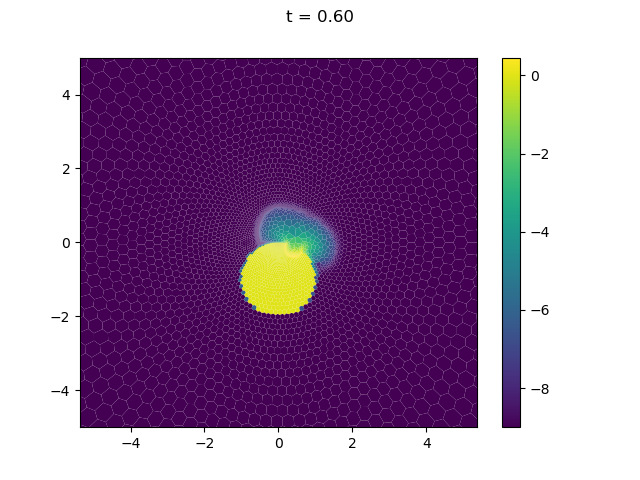}
\end{subfigure}
\begin{subfigure}{.49\textwidth}
  \centering
  \includegraphics[width=1\linewidth]{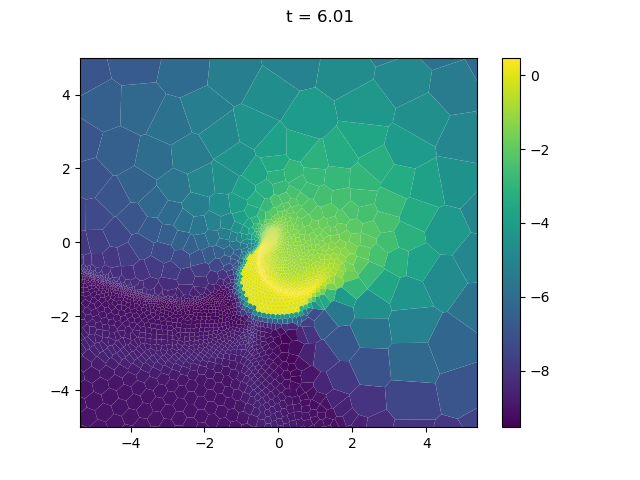}
\end{subfigure}
\begin{subfigure}{.49\textwidth}
  \centering
  \includegraphics[width=1\linewidth]{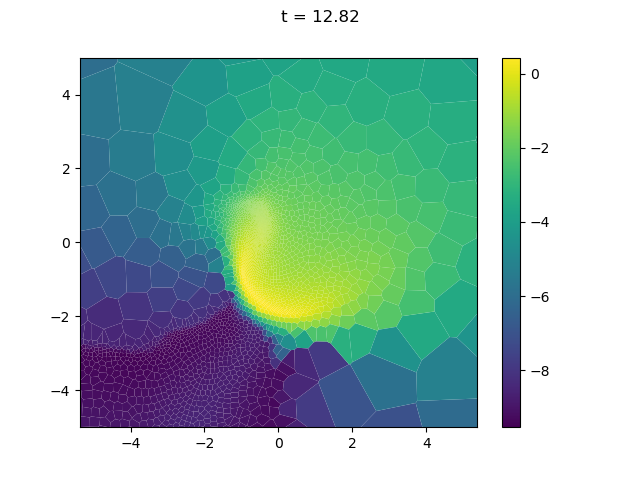}
\end{subfigure}

\caption{Snapshots from a simulation at different times of a collision between a planet and a planetary core whose radius is a factor of $0.1$ smaller. We do not include gravity, so the planet is obliterated in the end. The simulation is performed using a moving mesh hydrodynamic simulation, so the polygons are actual computational cells. \label{fig:sequence}
}
\end{figure}

Figure \ref{fig:sequence} shows four log density snapshots from an oblique collision, where the radius of the impactor is $0.1$ times the radius of the planet. The figure shows a shock wave emanating from the impact site that travels through the planet, and obliterates it, since there is not gravity to hold it together.

As the shock moves through the planet, it moves the ground, which pushes against the atmospheric gas column above it. In this sense, the ground acts like a piston, which sends a shock that moves up and away from the ground, accelerating due to the declining atmospheric mass density. The mass loss from each gas column as a function of ground velocity has been worked out in \cite{Schlichting2015AtmosphericImpacts}. By integrating the local atmospheric loss across the entire surface we can obtain the total atmospheric mass loss. By running the simulation with different impact parameter (or offset) and different impactor to target radius ratios (but fixed velocity, which we set to be 1.5 times the escape velocity), we obtain a map of atmospheric mass loss shown in Figure 6. One of the interesting features of this map is that for small impactors, the offset has very little influence on the outcome. This is because small enough impactors deposit all their energy in the target regardless of obliquity. This trend is in agreement with the findings of \cite{Yalinewich2018AtmosphericImpacts}, where it was shown that when the shock radius is considerably larger than the size of the impactor, the shock wave tends to the self similar solution found in the previous section, and in the process loses some of the information about the initial conditions (i.e. fast and small impactors create the same shock wave as slow and big impactors). The loss of information about the initial conditions is a common consequence of self similar solution \cite{Barenblatt1996ScalingAsymptotics}.

\begin{figure}
    \centering
    \includegraphics[width=1\linewidth]{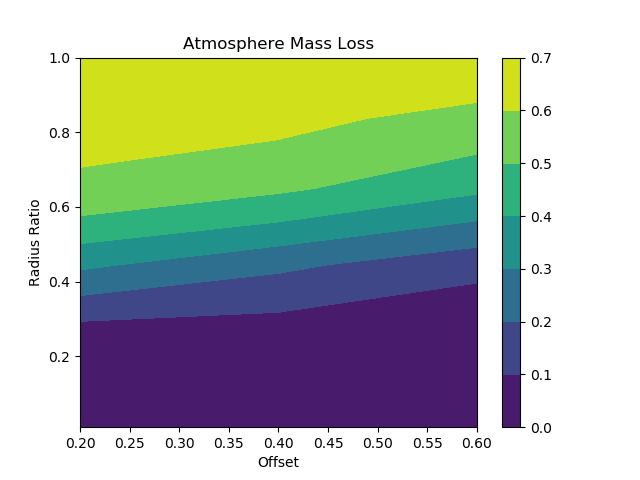}
    \caption{A map of the atmospheric mass loss as a function of the radius ratio between the impactor and target, and the offset between the centres normal to the initial velocity difference, or impact parameter. The velocity is held fixed at 1.5 of the escape velocity.}
    \label{fig:aml}
\end{figure}

\section{Conclusion} \label{sec:conclusion}
Planet formation involves a phase of giant collisions, when planets and planetary cores smash into each other. These collisions erode the planet's atmosphere. A key ingredient in modelling this mass loss is the evolution of the resulting shock wave in the interior of the planet. 

In order to justify the use of self similar solutions, we had to make a number of simplifying assumptions. First, we had to neglect all dimensional parameters in the problem, so such solutions only apply for shock radii much larger than the impactor on the one hand, but much smaller than the planet on the other hand. Second, we had to assume a strong shock, so that the impact velocity has to be much larger than the escape velocity. Third, we used an ideal gas equation of state, whereas real materials have a more complicated equation of state, which includes phase transitions. We argue that at high pressures and velocities, all materials behave like an ideal gases. Moreover, high pressure experiments can be used to determine the effective adiabatic index for different materials \cite{Yalinewich2018AtmosphericImpacts}.

Even with all these assumptions, the problem is still too complicated to be solved analytically, so we simplify it even further by considering the one dimensional analogue - the impulsive piston problem. This problem is a type II self similar shocks, where the evolution of the shock wave is not determined by some conservation law, but according to a singularity behind the shock front. By requiring that the hydrodynamic profiles satisfy both conditions at the shock front and the singularity, we can obtain $\delta = d \ln \dot{X} / d \ln m$, where $\dot{X}$ is the shock velocity and $m$ is the swept up mass. It turns out that this relation holds even in different geometries, i.e. in the three dimensional impact.

The impulsive piston problem has been originally solved by \cite{Adamskii1959TheLaw}. That work, however, used a different set of dimensionless variables, and so they could not reduce all equations to a single ordinary differential equation. They therefore had to use a multidimensional shooting method to determine $\delta$. The variables we chose allow us to reduce the problem to a single ordinary differential equation, but it comes at a cost. The cost is that we introduce a coordinate singularity, i.e. a place where the dimensional variables remain finite, but the dimensionless variables diverge. The price that we have to pay for this choice is that for some range of values for the adiabatic index $2 > \gamma >1$ the numerical integration is not continuous.

It is surprising that despite the many simplifying assumptions made here, the amounts of atmospheric mass loss predicted from this model is similar to more complicated models, which take into account all the effects considered here \cite{Yalinewich2018AtmosphericImpacts}. In the future, it would be interesting to see if this model can be refined to include some of the effects previously neglected.

\section*{Acknowledgements}
AY would like to thank Hilke Schlichting and Re'em Sari for the useful discussions. AY is supported by the  Vincent  and  Beatrice  Tremaine  Fellowship. This work made use of the sympy \cite{Meurer2017SymPy:Python}, numpy \cite{Oliphant2006ANumPy} and matplotlib \cite{Hunter2007Matplotlib:Environment} python packages.

\bibliographystyle{plain}
\bibliography{main}

\end{document}